\documentclass[aps,preprintnumbers,twocolumn,
nofootinbib
]{revtex4}  
\usepackage{graphicx}  
\usepackage{dcolumn}   
\usepackage{bm}        
\usepackage{amssymb}   
\usepackage{amsmath} 
\usepackage{comment}
\begin{document}

\widetext

\preprint{KOBE-COSMO-21-13, OCU-PHYS-540, AP-GR-168}



\title{Universal $10^{20}$\! Hz stochastic gravitational waves from photon spheres of black holes}


\author{Kaishu Saito$^1$}
\author{Jiro Soda$^1$}        
\author{Hirotaka Yoshino$^{2,1}$}

\affiliation{$^1$Department of Physics, Kobe University, Kobe 657-8501, Japan}  
\affiliation{$^2$Advanced Mathematical Institute, Osaka City University, Osaka 558-8585, Japan}


%
%
\begin{abstract}
We show that photon spheres of supermassive black holes generate high-frequency stochastic gravitational waves through
the photon-graviton conversion. 
Remarkably, the frequency is universally determined as $m_e\sqrt{m_e /m_p}  \simeq 10^{20} \text{Hz}$ in terms of the proton mass $m_p$ and the electron mass $m_e$.
 It turns out that the density parameter of the stochastic gravitational waves  $ \Omega_{ \text{gw}}$ could be  $ 10^{-12}$.
Since the existence of the gravitational waves from photon spheres is robust, 
it is worth seeking methods of  detecting high-frequency gravitational waves around $10^{20}$Hz.
\end{abstract}
\maketitle


%
%
\section{Introduction}
 The era of gravitational wave astronomy commenced~\cite{Arimoto:2021cwc} 
when  the first gravitational wave event from a binary black hole merger was detected~\cite{Abbott:2016blz}. 
Gravitational waves provide us with new information about the universe even before the recombination epoch which cannot be observed with electromagnetic waves.
However, the frequency range of gravitational waves that we have observed is rather narrow, $10~\text{Hz} \sim 10~\text{kHz}$, while that of electromagnetic waves has covered the frequency range
from kHz (radio) to $10^{26}~\mathrm{Hz}$ (gamma ray).  Apparently, it is important to extend the frequency frontier of gravitational wave observations.

There are many plans for observing gravitational waves from nHz to 100~MHz~\cite{Kuroda:2015owv}.
Beyond 100~MHz, however, the detection method of gravitational waves is different from that of gravitational wave interferometry~\cite{Aggarwal:2020olq}. For example, magnons in cavities are used to obtain the  sensitivity at $8~\text{GHz}$ and $14~\text{GHz}$~\cite{Ito:2019wcb}. 
Also detectors based on the photon-graviton conversion phenomenon~\cite{ME_Gertsenshtein} have been proposed to observe gravitational waves from planetary-mass primordial black hole binaries(typically $ 200~\text{MHz}$) \cite{Herman:2020wao}. 
The photon-graviton conversion is also used as a way to constrain the stochastic background gravitational waves with frequencies above $1~\text{THz}$ \cite{Ejlli:2019bqj}.

  In order to boost the study of high-frequency gravitational wave detectors, the existence of guaranteed  sources which emit high-frequency gravitational waves is essential. 
So far, various high-frequency gravitational wave sources have been proposed~\cite{Figueroa:2017vfa,Auclair:2019wcv,Dror:2019syi,Brustein:1995ki,Giovannini:1999bh,Wen et al.(2017),Ema:2020ggo,BisnovatyiKogan:2004bk,Ghiglieri:2015nfa,Aggarwal:2020olq,Fujita:2020rdx}.
In particular, primordial black holes (PBHs) \cite{Carr:1974nx}  is a possible source of high-frequency gravitational waves~\cite{Anantua:2008am,Dolgov:2011cq,Inomata:2020lmk,Dong:2015yjs}. 
PBHs evaporated before the big bang nucleosynthesis produce stochastic back ground around $10^{15}~\text{Hz} \sim 10^{ 19}~\text{Hz}$ with its typical density parameter $\Omega_{\text{gw}} \sim 10^{-7.5}$ . At present, however, it would be fair to say that  the existence of these sources are not guaranteed.

  In this paper, we propose a  novel source of high-frequency gravitational waves, namely, magnetospheres of supermassive black holes.
The stochastic gravitational waves  can be generated through the photon-graviton conversion from photon spheres of supermassive  black holes.
Indeed,  in the photon sphere of a black hole, the steady photon accretion from accretion disk effectively increases the conversion probability, and sufficient amount of gravitons are produced.
Since the existence of supermassive black holes has been already proved,  they are guaranteed sources.
We show that the frequency of these gravitational waves does not depend on the mass of black hole and strength of magnetic field under the assumption of equipartition of energies. 
Remarkably, the frequency is universally determined as $m_e\sqrt{m_e /m_p}  \simeq 10^{20}~\text{Hz}$ in terms of the proton mass $m_p$ and the electron mass $m_e$. 
We estimate contribution from super-massive black holes (SMBHs) to stochastic gravitational wave background
 and find that the density parameter of the stochastic gravitational waves  $ \Omega_{ \text{gw}}$ could be of the order of $ 10^{-12}$. It encourages us to seek detectors for high-frequency gravitational waves.

%
%
\section{Photon spheres of black holes}
In this section, we relate the photon intensity emitted from the accretion disk to the flux flowing into the vicinity of the photon sphere. These photons can stay in the black hole magnetosphere for a relatively long time and converted into gravitons, as will be explained later. 

\begin{figure}[tb]
  \centering
\includegraphics[width=0.95\linewidth]{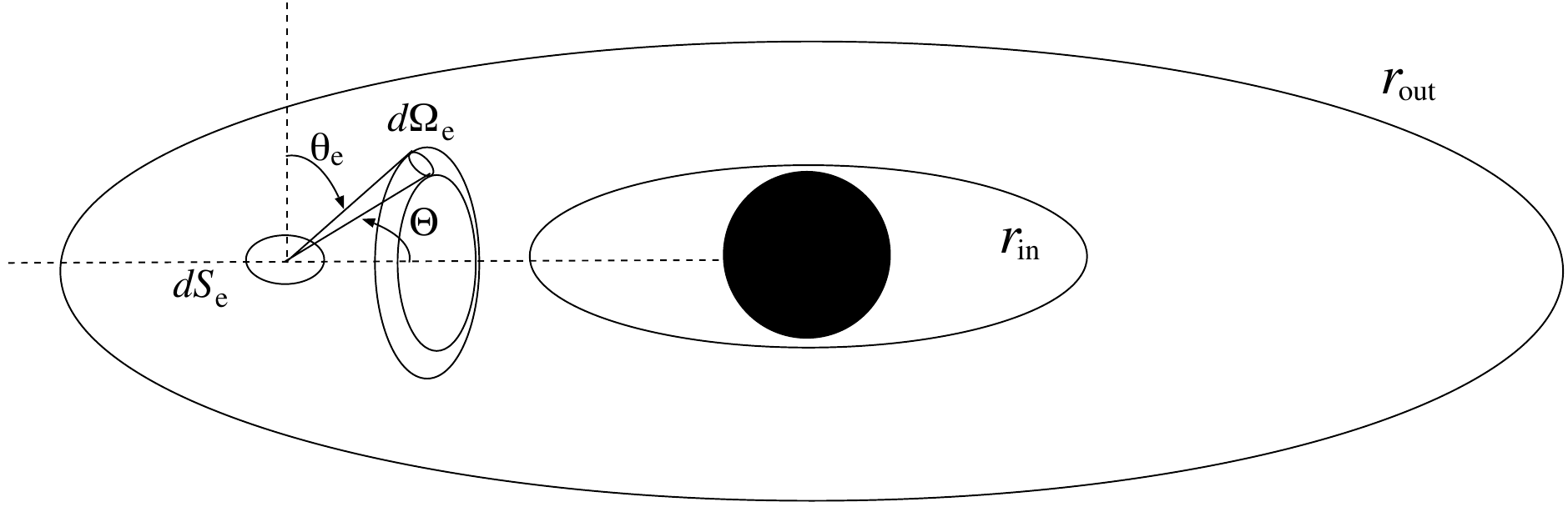}
\caption{Photon radiation from a thin accretion disk.}
\label{radiation}
\end{figure}

For simplicity, we assume the system to be a Schwarzschild black hole with a thin accretion disk.
The Schwarzschild metric is given by 
\begin{align}
ds^2 = - f(r) c^2 dt^2 + \frac{dr^2}{f(r)} + r^2 ( d \theta^2 + \sin^2 \theta d \phi^2 ),
\end{align}
with $f(r) = 1- r_{g}/r$ and gravitational radius $r_{g} :=2 GM/c^2$.
According to the Lambert's law, the number of photons emitted from an area element $dS_{e}$ of an accretion disk per unit time and passing through the solid angle $d \Omega_{e}$ is given by
\begin{align}
d^3  \left( \frac{d N}{d \tau_{e}} \right) = I^{(N)}_{e} ( \omega)  \cos \theta_{e} \,   d \Omega_{e} \, dS_{e} \, d \omega ,
\label{Lambert1}
\end{align}
where $d \tau_{e}$ is the proper time at the area element $dS_{e}$ located at the radius $r_{e}$, and $I^{(N)}_{e}(\omega)$ is a photon number intensity with an angular frequency $\omega$. For the angular coordinates used here, see Figure \ref{radiation}. We use subscript $e$ to refer to coordinates on the accretion disk.

%
%

  It is useful to introduce an impact parameter $b:=  \frac{r_{e}}{ \sqrt{f(r_{e})}} \sin \Theta \sim   \frac{r_{e}}{ \sqrt{f(r_{e})}} \sin(\frac{\pi}{2} - \theta_{e})$ instead of $\theta_{e}$ \cite{Yoshino:2019qsh}. The area element can be written as $dS_{e} = 2 \pi r_{e} d r_{e} / \sqrt{f(r_{e})}$. Then, we can rewrite Eq.~\eqref{Lambert1} as
\begin{align}
d^3 \left( \frac{dN}{dt} \right) =& (2 \pi)^2 I^{(N)}_{e} ( \omega) \left( \frac{f(r_{e})}{r_{e}} b^2 \right)  \notag \\
& \times  \frac{db}{ \sqrt{ {r_{e}}^2 /f(r_{e}) -b^2 }} dr_{e}  \, d \omega ,
\end{align}
where $t$ is the Schwarzschild time. 
To evaluate this, we replace $b$ with the critical impact parameter $b_{\text{crit}}=3 \sqrt{3}  \, r_{g}/2$ which correspond to circular orbit.
Integrating radial coordinate $r_{e}$ from $3 r_{g}$ to infinity, we obtain the photon number emitted to $b =b_{\text{crit}} + db$ per unit time and unit angular frequency
\begin{align}
d \left( \frac{d^2 N}{dt \, d \omega} \right) \sim 27 \, r_{g} {I_{e}}^{(N)}( \omega) db .
\label{photon_flux_to_b}
\end{align}
This equation is useful in Sec.~IV to estimate the number of gravitons produced around the photon sphere.

%
%
\section{Photon-Graviton Conversion}
In this section we shall briefly review the conversion phenomenon between photon and graviton proposed by Gertsenshtein \cite{ME_Gertsenshtein} with which we shall explore the possibility of gravitational wave emission from black hole magnetospheres. For more detailed derivation, see \cite{Masaki:2018eut}.

We consider the Einstein-Hilbert action with the Euler-Heisenberg effective Lagrangian for electromagnetic fields minimally coupled to the gravity,
\begin{align}
S &=  \int d^4 x \sqrt{-g} \left[ \frac{1}{16 \pi G} R       
   - \frac{1}{4}  F_{\mu \nu} F^{\mu \nu}  \right]          \notag\\
&+  \frac{ \alpha^2}{90m_{e}^4}\int d^4 x \sqrt{-g} \left[  (F_{\mu \nu} F^{\mu \nu})^2   + \frac{7}{4} (\tilde{F}_{\mu \nu} F^{\mu \nu})^2   \right],
\label{EH+EH}
\end{align}
where $\alpha=7.297...\times10^{-3}$ is the fine structure constant and $m_{e}$ is the electron mass. The field strength is defined by $F_{\mu \nu} := \partial_{\mu} A_{\nu} - \partial_{\nu} A_{\mu}$.
 In order to describe propagations of gravitons and photons, we consider perturbations around background fields,
\begin{align}
 g_{\mu \nu} = \eta_{\mu \nu} + \kappa h_{\mu \nu} \ , \quad 
 A_{\mu} = \bar{A}_{\mu} + \cal{A}_{\mu},
\end{align}
 where $\eta_{\mu \nu}$ is the Minkowski metric and $\kappa := \sqrt{16 \pi G}$.
 Hereafter, we impose the TT gauge for $h_{\mu \nu}$ and the radiation gauge for $A_{\mu}$.
 The background magnetic field is assumed to be static and uniform and to be aligned in the $y$ direction of the Cartesian coordinates. 
 
   The conversion phenomenon occurs when gravitational waves propagate in the $z$ direction, i.e., a direction perpendicular to the magnetic field. The plane-wave configuration along the $z$ direction are expanded as 
 \begin{align}
 {\cal{A}}_{i} (z,t) &= i {\cal{A}}_{+}(z) u_{i} e^{- i \omega t} + i {\cal{A}}_{\times}(z) v_{i} e^{-i \omega t},  \\
 h_{ij}(z,t) &= h_{+}(z) e_{ij}^{+} e^{-i \omega t} + h_{\times}(z) e^{\times}_{ij} e^{-i \omega t}, 
 \end{align}
 where $u_{i},v_{i},e^{+}_{ij},e^{\times}_{ij}$ are the polarization
 vectors and tensors.
  Substituting this plane-wave configuration for the linearized equation of motion obtained from Eq.~\eqref{EH+EH}, one gets a Schr\"{o}dinger-type equation
 \begin{align}
 i \frac{d}{dz} \psi(z) = {\cal{M}}  \psi(z),
 \label{sch_eq}
 \end{align}
 where
 \begin{align}
 \psi (z) := 
 \begin{pmatrix}
 h_{\lambda}(z) \\
 {\cal{A}}_{\lambda}(z) \\
 \end{pmatrix}
 e^{-i \omega z} ,
\,\,\,\,\,  
\cal{M} =
\begin{pmatrix}
0                             &               \Delta_{g \gamma}    \\
\Delta_{g \gamma} &               \Delta_{\gamma}       \\
\end{pmatrix}
 \end{align}
 and $\lambda$ stands for the polarization, $\lambda = +,  \times$.
 Here,  $\cal{M}$ is called a mixing matrix which describes the effective photon mass, $\Delta_{\gamma}$, and the coupling between gravitons and photons, $\Delta_{g \gamma} := 2 \sqrt{ \pi} B/ M_{\text{pl}}$.
$\Delta_{\gamma}$ consists of two parts as
\begin{align}
\Delta_{\gamma} =  \Delta_{\rm QED} + \Delta_{p} .
\end{align}
Here, $\Delta_{\rm QED}$ describes the effective
photon mass that originates from the effect of
the quantum electrodynamics (QED).
The QED contribution comes from the Euler-Heisenberg
effective action in Eq.~\eqref{EH+EH},
and depends on the polarization $\lambda$ as
\begin{align}
\Delta_{\text{QED} \lambda } = - k(\lambda) \varrho \omega B^2 , \qquad \varrho = \frac{4 \alpha^2}{45 m_{e}^4},
\end{align}
where $k(+) = 2, \,\, k(\times) = 7/2$.
The term $\Delta_{p}$ 
originates from the fact
that in the presence of plasma,
electromagnetic waves have an effective mass that corresponds to
the plasma angular frequency $\omega_p$.
It is given as
\begin{align}
\Delta_{p} = \frac{ \omega^2_{p}}{2 \omega} \quad \text{and} \quad  \omega_{p}^2 = 4 \pi \alpha \frac{n_{e}}{m_{e}}
\end {align}
with the electron number density $n_{e}$.
Since the mixing matrix $\cal{M}$ is constant, under the initial condition $h_{\lambda}(0) =1 , {\cal{A}}_{\lambda}(0)=0$,  Eq.~\eqref{sch_eq} is easily solved as 
\begin{align}
h_{\lambda} &= \cos^2 \theta e^{- i \lambda_{+} z} + \sin^2 \theta e^{- i \lambda_{-} z} , \notag \\
{\cal{A}}_{\lambda} (z) &= \cos \theta \sin \theta e^{-i \lambda_{+} z} - \cos \theta \sin \theta e^{-i \lambda_{-} z},
\end{align}
where $\theta$ is the mixing angle given by 
\begin{align}
& \cos 2 \theta = - \frac{ \Delta_{\gamma} }{ \Delta_{\text{osc}}} , \quad \sin 2 \theta = \frac{2 \Delta_{g \gamma}}{\Delta _{\text{osc}}} ,   \notag  \\
&  \text{and}  \quad   \Delta_{\text{osc}} := \sqrt{ \Delta_{ \gamma}^2 + (2 \Delta_{g \gamma})^2 }.
\end{align}
Here, we defined the eigenvalues of $\cal{M}$ by $\lambda_{\pm}$.
Therefore, the conversion probability from gravitons to photons after propagating the distance $z$ becomes
\begin{align}
P  &= \left(  \frac{2  \Delta_{g \gamma}}{ \Delta_{ \text{osc}}} \right)^2 \sin^2 \left( \frac{ \Delta_{ \text{osc}}}{2} z  \right).  \label{conv_prob}  
%
%
\end{align} 
The complete conversion is possible only when the coefficient $ \left( 2 \Delta_{g \gamma} / \Delta_{ \text{osc}} \right)^2$ becomes unity. In other words, it is only possible at the resonance frequency 
where the effective photon mass vanishes. By solving $\Delta_{\gamma} =0$, the resonance angular frequency is given by 
\begin{align}
\omega_{r}^2 = \frac{\omega_{p}^2}{2} \frac{1}{k \varrho B^2} = \frac{45 \pi}{2 k \alpha} \frac{n_{e} m_{e}^3}{B^2}.
\label{resonance_freq}
\end{align}
This is completely determined by $k$, the magnetic field $B$ and the plasma density $n_{e}$. 
%
At the resonance frequency, conversion phenomenon occurs when a phase of the conversion probability of Eq.~\eqref{conv_prob} become $\pi /2$.
This determines the conversion length as
\begin{align}
L_{\gamma \leftrightarrow g } &=1.7  \left( \frac{10^{12} \text{G}}{B}  \right) 10^{-6} \text{pc}  \notag \\
&= 5.4  \left( \frac{10^{12} \text{G}}{B}  \right) 10^{10}  \text{m}.
\end{align}
The fact that this conversion length is very long is the reason why the photon-graviton 
conversion rarely occurs. However, as we will see in the next section,  the conversion occurs effectively in the vicinity of a photon sphere of a black hole.

Let us estimate the angular frequency range that contributes to the conversion phenomenon.
First, the conversion phenomenon occurs sufficiently (i.e. $P\sim 1$) even at angular frequency $\omega = \omega_{r} +\Delta \omega_{r}$, which is close to the resonance angular frequency. Then, let us estimate the order of $\Delta \omega_{r}$.
Since the factor of $ \left( 2 \Delta_{g \gamma} / \Delta_{ \text{osc}} \right)^2$
in Eq.~\eqref{conv_prob} is written as
$1/\left[1+( {\Delta_{\gamma}}/{2 \Delta_{g \gamma}})^{2}\right]$,
the angular frequency $\omega$ should satisfy 
\begin{align}
\Delta_{\gamma} \lesssim 2 \Delta_{g \gamma}  \ 
\end{align}
for the conversion probability to become $O(1)$. 
In this inequality, $\Delta_{\gamma}$ depends on $\omega = \omega_{r} +\Delta \omega_{r} $. Expanding around $\omega_{r}$, we obtain the following inequality for $\Delta \omega_{r}$:
\begin{align}
\Delta \omega_{r} \lesssim   \frac{2 \sqrt{\pi}}{k} \frac{1}{M_{\text{pl}} \varrho B}   = \frac{45 \sqrt{\pi}}{2 k \alpha^2} \frac{m_{e}^4}{M_{\text{pl}} B}.
\end{align}
If we use $k\sim 3 $ this becomes
\begin{align}
\Delta \omega_{r} \lesssim  0.36 \left( \frac{10^{12} \text{G} }{B} \right) \text{MHz}.
\label{freq_band}
\end{align}
This upper bound determines the frequency band where the conversion occurs sufficiently, and depends only on the magnetic field $B$.

%
%
\section{Universal GRAVITATIONAL WAVES from photon spheres}
In the following, we will apply the photon-graviton conversion phenomena discussed in the previous section to the system of a single black hole with mass $M$ and the accretion disk around it and show that 
gravitational waves with the universal frequency are emitted from the vicinity of the photon sphere.
The idea is quite simple: photons from the accretion disk steadily accrete around the photon sphere, and those with the resonance angular frequency $\omega_{r}$ are converted into gravitons with the same angular frequency $\omega_{r}$ by the magnetic field of the magnetosphere. 

In an inner disk, we assume the thermal population of electrons. 
Then, the equipartition of energy
  between the plasma and the magnetic fields is expected be satisfied \cite{Shvartsman,Marrone:2005ky,Wallace:2021ncq},
\begin{align}
\frac{B^2}{8 \pi} \sim m_{p} c^2 n_{e} ,
\label{equipartition}
\end{align}
where $m_{p}$ is the proton mass. Surprisingly, under the assumption of the equipartition of energy, 
the resonance angular frequency of Eq.~\eqref{resonance_freq} is independent of the magnetic field $B$ and the plasma density $n_{e}$, and it is estimated as 
\begin{align}
\omega_{r} \sim \left( \frac{45}{16 k \alpha} \frac{m_{e}^3}{m_{p} c^2} \right)^{1/2} \sim 2.1 \times 10^{20} \,  \text{Hz} .
\end{align}
Therefore, the typical frequency of gravitational waves emitted from the magnetosphere is $10^{20}~\text{Hz}$, regardless of the details of the black holes such as the mass $M$ and the magnetic field $B$. This is a very robust and universal result.

   Let us estimate the luminosity of gravitational waves by counting the number of gravitons produced by the conversion. First, we focus on a given impact parameter $b$ around $b_{\text{crit}}$.
 A photon with the impact parameter $b$ will stay around the photon sphere for a period $T(b) := - \frac{3 \sqrt{3}}{2c} r_{g} \log | 2(b-b_{\text{crit}})/r_{g}| $ \cite{Yoshino:2019qsh}. Therefore, for a photon with the impact parameter $b$, the conversion probability from a photon to a graviton is given by $P(cT(b))$.
 Taking into account the photon flux with the impact parameter $b$ given by Eq.~\eqref{photon_flux_to_b}, the number of produced gravitons is given by
 \begin{align}
 d \left( \frac{d^2 N_{\gamma \to g}}{dt \, d \omega} \right) \sim 27 \, r_{g} {I_{e}}^{(N)}( \omega_{r}) db \times P(cT(b)) .
 \end{align}
 Integrating the above quantity with respect to  $b \in \left[  b_{\text{crit}} -r_{g}/2, b_{\text{crit}} +r_{g} /2  \right]$
 \footnote{Here, we used the analytic formula
 $
 \int_{-1}^{1} \sin^2  \left( a \log |y| \right) d y = \frac{(2a)^2}{1+ (2a)^2}.
 $
 }
 , we obtain the number of gravitons produced in the vicinity of the photon sphere per unit time,
\begin{align}
\frac{d N_{ \gamma \to g}}{dt} = \frac{27}{2} \,  \frac{27 (r_{g} \Delta_{g \gamma})^2}{1 + 27 ( r_{g} \Delta_{g \gamma})^2} \, r_{g}^2 {I_{e}}^{(N)}( \omega_{r}) \Delta \omega_{r}.
\label{graviton_number_1}
\end{align}
Here, $r_{g} \Delta_{g \gamma}$ is a useful dimensionless parameter given by
\begin{align}
r_{g} \Delta_{g \gamma} = 2.59 \times 10^{-20} \left( \frac{M}{ M_{\odot}} \right) \left( \frac{B}{1 \text{G}} \right).
\end{align}
As long as the equipartition principle holds, this parameter is very small compared to unity. 
Therefore, we can approximate Eq.~\eqref{graviton_number_1} as 
\begin{align}
\frac{d N_{ \gamma \to g}}{dt} = \frac{729}{2} \xi \,  \left( r_{g} \Delta_{g \gamma}  \right)^2  L(\omega_{r}) \frac{ \Delta \omega_{r}}{ \hbar \omega_{r}}.
\label{graviton_number_2}
\end{align}
Here, we introduced a spectral photon luminosity $L( \omega_{r})$ which represents the photon energy flux from the accretion disk per unit time and unit angular frequency. The dimensionless factor $\xi$ is defined by $\xi:= r_{g}^2/ A( r_{g})$ with the area of the accretion disk $A(r_{g})$ such that it emits X-rays near the resonance frequency $\sim 10^{20} \text{Hz}$.
Multiplying Eq.~\eqref{graviton_number_2} by the energy of a graviton,  we obtain the luminosity of gravitational waves as
\begin{align}
\frac{d E_{ \gamma \to g}}{dt} = \frac{729}{2} \xi \,  \left( r_{g} \Delta_{g \gamma}  \right)^2  L(\omega_{r}) \Delta \omega_{r} \ . 
\label{gw_luminnosity}
\end{align}
To evaluate this luminosity, we use a band given by Eq.~\eqref{freq_band} where the conversion sufficiently occurs; $\Delta \omega_{r}:= 0.36 \left( \frac{10^{12} \text{G}}{B} \right) \text{MHz} $. According to the deep X-ray survey data from Chandra observatory \cite{Georgakakis:2015rfa}, AGN X-ray luminosity is typically $L_{X}(2-10 \text{keV}) \sim 10^{43} - 10^{46} \text{erg} \text{sec}^{-1}$.
This roughly corresponds to the Eddington luminosity of a supermassive black holes. For this reason, we will assume that the luminosity of the X-ray band is the Eddington luminosity. Thus, the spectral luminosity of a typical AGN reads
 \begin{multline}
 L( \omega_{r} \sim 10^{20}~\text{Hz}) \ \sim \ \frac{L_{\text{Edd}}(M)}{10^{20}~\text{Hz}}  \\
 \sim \ 10^{24} \left(\frac{M}{10^{6} M_{\odot}}  \right) \text{erg} \, \text{sec}^{-1} \, \text{Hz}^{-1} ,
\end{multline}
where $M$ is the mass of the central black hole.
Finally,  we obtain the typical luminosity of  gravitational waves emitted from the photon sphere of a single black hole 
\begin{align}
\frac{dE}{dt} \sim 8.80 \xi  \times 10^{22} \left( \frac{B}{10^6 \text{G}} \right)     \left( \frac{M}{10^6 M_{\odot}} \right)^3 \text{erg} \, \text{sec}^{-1}.
\end{align}

%
%
\section{high-frequency stochastic GRAVITATIONAL WAVES }
As we saw in the previous section, the frequency of gravitational waves from a photon sphere of a black hole is universally given by  $ 10^{20}~\text{Hz}$  which dose not depend on the details of black holes.
Apparently, the luminosity of gravitational waves emitted by each black hole is rather faint.
Nevertheless, we can observe them as background gravitational waves. In the following, we will calculate the density parameter of the stochastic gravitational waves.

  For simplicity, we ignore the effects of the expansion of the universe and assume the equipartition of the energy between the  energy of magnetic fields and the Eddington luminosity.
Then, we obtain
\begin{eqnarray}
\frac{B^2}{8 \pi} 4 \pi \left( \frac{3}{2} r_{g} \right)^2 c = L_{ \text{Edd} }( M)  \ .
\end{eqnarray}
Thus, the magnetic field can be estimated as
\begin{eqnarray}
 B \sim 0.241 \times  10^{7} \sqrt{ \frac{10^{6} M_{\odot}}{M} } \ \text{G} \  .
\end{eqnarray}
Under this assumption, the luminosity of gravitational waves from a single black hole 
can be written as $\frac{dE}{dt} ( B( M),M)$.
 We set our position at the origin and consider the radiation from a point $\bm{x}$. Assume that there are $n( \bm{x},M) d^3  \bm{x} \, dM$ black holes of mass $M$ at the point $\bm{x}$. Since the luminosity $dE/dt (B(M ),M)$ is conserved on the sphere surrounding the point $\bm{x} $, 
 the energy density of gravitational waves we observe reads
\begin{align}
\rho_{\text{gw}}( \bm{x} ,M) = \frac{1}{4 \pi c} d^3 \bm{x} \, dM \frac{1}{| \bm{x}|^2} n( \bm{x},M) \frac{dE}{dt}(B(M),M).
\end{align}
Here, the arguments of the left-hand side imply that $\rho_{\text{gw}}( \bm{x},M)$ is contribution from a point $\bm{x}$ and a given mass $M$. Integrating this over the entire observable region and the mass of the black holes and dividing it by the critical energy density of the universe, 
we obtain the density parameter of the stochastic gravitational waves as
\begin{multline}
{h_{0}}^2 \Omega_{\text{gw}} \ =\  \frac{2G {h_{0}}^2}{3c^3 H_{0}^2}  
\\
\times \int dM \frac{dE}{dt}(B(M),M) \, \int_{| \bm{x} | \le c {H_{0}}^{-1} } d^3 \bm{x} \frac{n( \bm{x},M)}{| \bm{x} |^2} \ ,
\label{STC_intensity}
\end{multline}
where $h_{0}$ is the dimensionless Hubble constant.

From the recent observations of Refs.~\cite{Christopher A Onken,Dullo,Mehrgan}, we see that the mass range of supermassive black holes is $10^{6} M_{\odot} \sim 10^{11} M_{\odot}$. 
It is natural to assume the distribution of black holes to be spatially homogeneous  and to take the power law with respect to the mass, 
\begin{align}
n(\bm{x},M) =  C( \beta) \frac{N_{\text{galaxy}}}{ \frac{4\pi}{3} (c H_{0}^{-1})^3 } M^{- \beta},
\end{align}
where $\beta \ge 0$ and $N_{\text{galaxy}} = 2 \times 10^{12} $ is the total number of galaxies.
The normalization constant $C( \beta)$ is determined by $\int_{10^{6} M_{\odot}}^{10^{11} M_{\odot}} dM \int d^{3} \bm{x} n( \bm{x},M) = N_{ \text{galaxy}}$
as
%
\begin{align}
C( \beta ) = \left( \int_{1}^{10^5} d y \,  y^{- \beta}  \right)^{-1} ( 10^6 M_{\odot})^{ \beta-1} \ .
\end{align}
Then, the density parameter \eqref{STC_intensity} can be easily integrated as 
\begin{align}
h_{0}^2 \Omega_{ \text{gw}} \sim 0.23 h_{0}^2  \xi  \times 10^{-23}  D( \beta),
\label{STC_intensity2}
\end{align}
where $\xi$ is the dimensionless ratio, $\xi := r_{g}^2/ A(r_{g})$, and $D( \beta)$ is defined by 
\begin{align}
D( \beta) = \frac{\int_{1}^{10^5} d y' \, {y'}^{5/2- \beta} }{\int_{1}^{10^5} d y \,  y^{- \beta} } .
\label{function_D}
\end{align}
%
\begin{figure}[h]
  \centering
\includegraphics[width=0.95\linewidth]{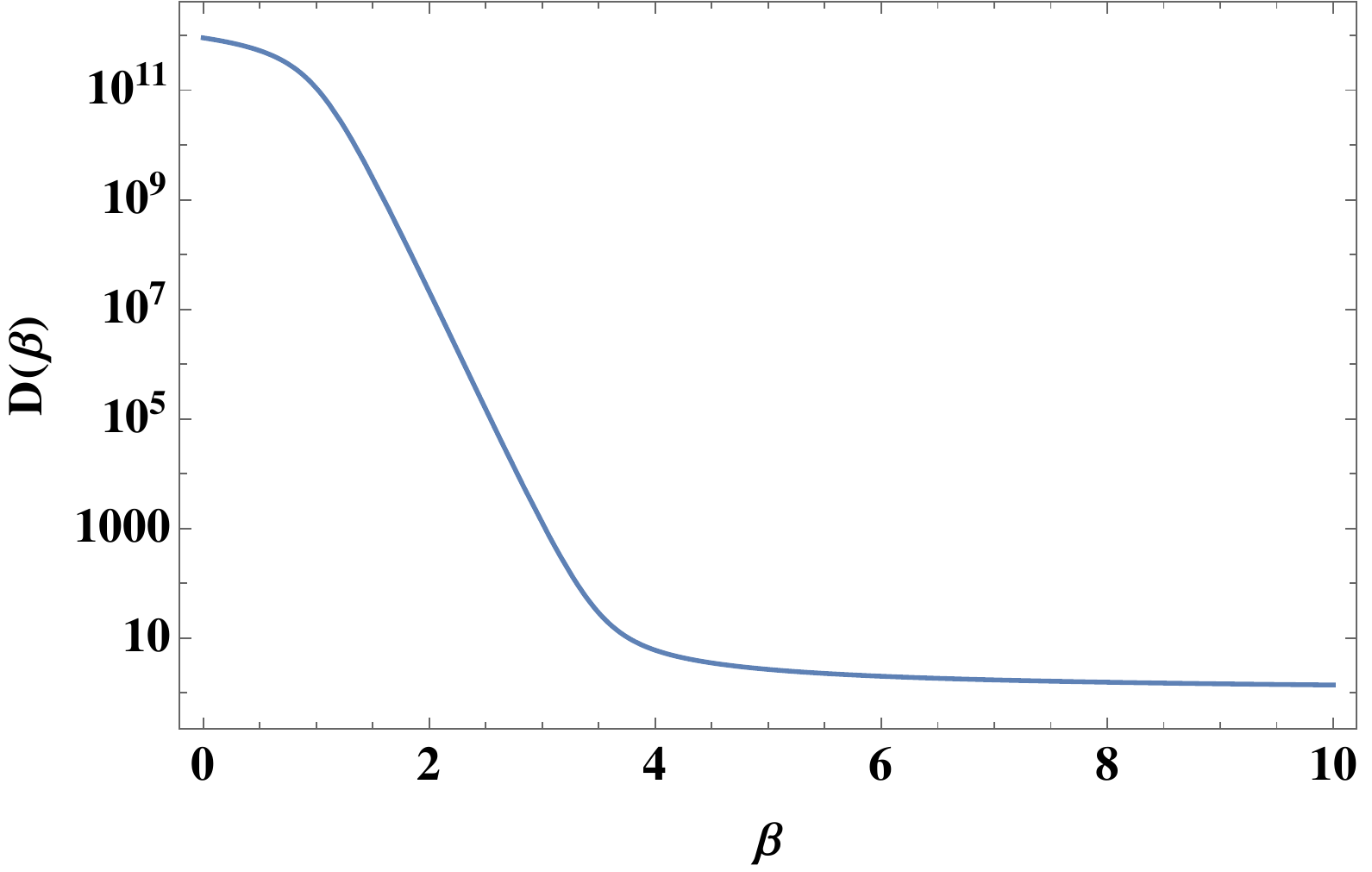}
\caption{Plot of $D(\beta)$.}
\label{D_Plot}
\end{figure}
%
From Fig.~\ref{D_Plot}, the energy density strongly depends on $\beta$ and monotonically decreases as the value of $\beta$ is increased.
Now let us estimate the maximum value of the energy density. Taking $\beta $ as $ 0 \le \beta \ll1$, the value of the constant $D(\beta)$ becomes
$D(\beta)\approx\frac{2(1- \beta)}{7-2\beta} \times 10^{25/2}$.
Thus, we finally obtain
\begin{align}
h_{0}^2 \Omega_{\text{gw}} = h_{0}^2 \xi \, \frac{2 (1- \beta)}{7 -2 \beta} \, 7.4 \times 10^{-12} .
\end{align}
Therefore, the order of $h_{0}^2 \Omega_{\text{gw}}$
could be as large as $10^{-12}$, and 
this gives rise to a motivation to invent detectors for high-frequency gravitational waves of the range around $10^{20}~\mathrm{Hz}$.

%
%
\section{conclusion}
In this paper, we proposed a novel and robust source of high-frequency gravitational waves. We have shown
that the photon spheres of  supermassive black holes emit gravitational waves through photon-graviton conversion phenomenon,
with the universal frequency $10^{20}~\text{Hz}$
which depends on proton and electron masses, not on black hole mass and its magnetic field. 
Such gravitational waves can be observed as the stochastic gravitational wave background. 
We estimated the density parameter of the stochastic gravitational waves and found that it could be as large as $h_{0}^2 \Omega_{ \text{gw}} \sim   10^{-12}$.

  As is evident from the formula of Eq.~\eqref{STC_intensity}, the density parameter depends on the black hole number density $n(\bm{x},M)$. Therefore, future observations of gravitational waves from black hole photon spheres may give a bound on $n( \bm{x},M)$. This bound may have a significant implication for the abundance of intermediate mass black holes, the number of which is currently unknown \cite{Kawaguchi:2007fz}.
  
  Another important point of our work is that our prediction is based only on Einstein gravity and Maxwell electrodynamics, which have been verified to a high degree of accuracy. 
Therefore, high-frequency gravitational waves predicted in this paper is robust. We hope that our result motivates and boosts the development of high-frequency gravitational wave detectors for the range around $10^{20}~\mathrm{Hz}$, and 
  opens a new window to explore the universe with gravitational waves.

%
%

\acknowledgments

J.\,S.  was in part supported by JSPS KAKENHI Grant Numbers JP17H02894, JP17K18778, JP20H01902.
H.\,Y. was in part supported by JSPS KAKENHI Grant Numbers JP17H02894,
JP18K03654. 
The work of H.Y. is partly supported by
Osaka City University Advanced Mathematical Institute
(MEXT Joint Usage/Research Center on Mathematics and Theoretical Physics 
JPMXP0619217849).


\end{document}